\documentclass{article}
\usepackage{times}

\usepackage{microtype}
\usepackage{graphicx}
\usepackage{booktabs}
\usepackage{multirow}
\usepackage{makecell}
\usepackage{tabularx}
\usepackage{url}
\usepackage[utf8]{inputenc}
\usepackage[T1]{fontenc}
\usepackage{float}
\usepackage{wrapfig}
\usepackage{placeins}
\usepackage{nicefrac}

\usepackage{hyperref}
\usepackage{xurl}

\usepackage[most]{tcolorbox}
\usepackage{amsmath}
\usepackage{amsfonts}
\usepackage{amssymb}
\usepackage{xcolor}
\usepackage{colortbl}

\usepackage{eso-pic}
\usepackage{setspace}
\usepackage{fancyhdr}
\usepackage{caption}
\usepackage{needspace}

\usepackage{nebius}
\hypersetup{pdftitle={Interpreting Coding Agents Through Their Behavioral Trajectories}, pdfauthor={Ibragim Badertdinov, Alexander Golubev}}

\graphicspath{{./}{figures/}}

\title{Interpreting Coding Agents Through Their Behavioral Trajectories}
\author{\textbf{Ibragim Badertdinov}\textsuperscript{1*} \and \textbf{Alexander Golubev}\textsuperscript{1}}
\date{}

\makeatletter
\renewcommand{\@maketitle}{%
  \vspace{-1pt}%
  {\centering\fontsize{18pt}{22pt}\selectfont\bfseries \@title\par}%
  \vspace{1.2em}%
  {\centering\renewcommand{\and}{\unskip,\ }%
   \begin{spacing}{1.15}\@author\end{spacing}\par}%
}
\makeatother

\renewenvironment{abstract}{\vskip.075in\centerline{\large\bfseries
Abstract}\vspace{0.5ex}\begin{quote}}{\par\end{quote}\vskip 1ex}

\AddToShipoutPicture*{%
  \AtTextUpperLeft{%
    \raisebox{5pt}{%
      \includegraphics[width=80px]{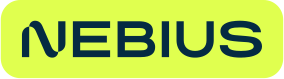}
    }%
  }%
}

\definecolor{bgcolor}{HTML}{EFF4F9}

\makeatletter
\@ifundefined{lx@tag}{%
  \newtcolorbox{roundedbox}{
    enhanced,
    colback=bgcolor,
    frame hidden,
    arc=3mm,
    auto outer arc,
    left=4pt, right=4pt,
    top=6pt, bottom=6pt,
  }%
}{%
  \newenvironment{roundedbox}{\par}{\par}%
}
\makeatother

\begin{document}
\onecolumn

\makeatletter
\setlength{\@fptop}{0pt}
\setlength{\@dblfptop}{0pt}
\makeatother

\thispagestyle{empty}

\begin{roundedbox}
  \centering

  \maketitle
  \thispagestyle{empty}
  \vspace{0.5em}

  {\small
  \textsuperscript{1}Nebius\\
  \vspace{0.3em}
  \textsuperscript{*}Correspondence to: ibragim-bad@nebius.com\\
  }
  \vspace{0.8em}

  \begin{abstract}
  Standard coding-agent benchmarks primarily evaluate binary task success, treating execution as a black box and obscuring the behaviors that produce those outcomes. We analyze a large corpus of 143,209 execution trajectories from 125 agent configurations spanning 117 unique models. To enable comparisons across heterogeneous agent interfaces, we develop a rule-based parser that maps raw execution traces to high-level actions such as repository search, file editing, and test execution, and audit its predictions against a six-model consensus reference on 500 events. We find that agents with similar solve rates can nevertheless differ substantially in how they explore, modify and verify code. Moreover, without using task outcomes, token counts, or tool names as features, aggregate behavioral profiles retrieve the corresponding agent stack across disjoint repository sets and programming languages, although task composition is not fully controlled in the cross-repository setting. Through analyses of frontier models and agent scaffolds, we further show that scaffold choice and reasoning effort systematically shift agents' search, editing, and testing behavior. We release the deterministic parser together with the full annotation and analysis pipeline alongside an annotated trajectory subset.
  \end{abstract}
\end{roundedbox}

\section{Introduction}

Coding-agent benchmarks typically reduce a complex, multi-step interaction to a binary outcome: whether the agent resolves the issue and passes the tests~\citep{jimenez2024swebench}. Aggregated across tasks, this produces a solve rate that is useful for ranking systems but offers limited insight into how they operate. The differences in behavior matter for diagnosing failures and understanding how different models approach the problems. Moreover, scaffolding can substantially alter both workflow and success rates~\citep{yang2024sweagent}, showing that behavior depends on the entire system, not just the underlying model.

Analyzing these behaviors at scale is challenging. Execution traces are heterogeneous: harnesses expose different tools, use different command formats, and record interactions under different schemas. Direct comparisons based on tool names or raw commands therefore conflate semantic actions with implementation details. A cross-system analysis requires a shared representation that captures what an agent is doing, independently of how a particular harness records that action.

In this work, we develop such a representation and apply it to 143,209 execution trajectories from SWE-rebench~\citep{badertdinov2025swerebench}, covering 125 agent configurations and 117 unique models. We introduce a rule-based parser that maps heterogeneous trace events to a common taxonomy of high-level actions. We audit the parser on a sample of 500 events and use the resulting annotations to compare agent workflows at scale. 

Our analysis shows that solve rate does not uniquely characterize agent behavior. 
Configurations with similar performance can differ substantially in how often they search, inspect, edit, and test. We further find that aggregate behavioral profiles contain enough information to predict the underlying model and harness on held-out repositories, even without task outcomes, token counts, or tool names. Finally, we examine frontier models and agent scaffolds to show how scaffold choice and reasoning effort systematically shift patterns of exploration, modification, and verification.

\needspace{6\baselineskip}
Our main contributions are as follows:
\begin{itemize}
    \item We introduce a deterministic trajectory normalization pipeline, annotate 143,209 trajectories from 117 models, and release the deterministic parser, analysis pipeline, and an annotated trajectory subset.
    \item Through outcome-matched comparisons, we show that nearly identical solve rates can conceal large and reproducible differences in observable workflows.
    \item We show that, without using outcomes, token counts, or native tool names as retrieval features, aggregate behavioral profiles can retrieve the deployed stack across disjoint repository sets and programming languages; task composition across repository splits remains a potential confound.
    \item We use the taxonomy to characterize how model choice, scaffolding, and reasoning effort affect agents' exploration, modification, and verification patterns.
\end{itemize}
\section{Related Work}

Recent efforts in software engineering benchmarks have expanded execution-based evaluation toward multi-dimensional assessments~\citep{kapoor2025aiagents} and contamination-resistant environments~\citep{badertdinov2025swerebench}. Because repeated rollouts on identical issues exhibit substantial behavioral divergence~\citep{bjarnason2026randomness}, understanding the underlying execution trajectory is essential to explain how different models and harnesses actually resolve code tasks.

Process-level analysis of coding agents investigates intermediate execution patterns beyond final pass rates. Prior work examines intermediate task progress~\citep{ma2024agentboard}, stage-based workflow labeling~\citep{sahoo2026agentlens}, and human-validated behavioral vocabularies~\citep{gao2026interpretagentbehavior}. Other approaches represent execution rollouts via process-flow graphs~\citep{liu2026processcentric} or programmatic procedural traces~\citep{oderinwale2026trajectories}. Recent studies also highlight how task composition influences behavior~\citep{mehtiyev2026beyond} and analyze specific stage bottlenecks across code search, reading, and editing~\citep{shu2026resolverate,kim2026coherencecollapse,bouzenia2025understanding,majgaonkar2025understanding,chen2026beyondfinal,ceka2026traceability}.

Trajectory failure analysis has increasingly moved toward temporal and structural breakdowns. \citet{zhao2026failureprocess} model error accumulation and recovery as dynamic processes, while TrajAudit uses reference-aware signals to identify exact failure points~\citep{wang2026trajaudit}. Furthermore, \citet{raj2026modelharness} emphasize that agent capabilities are tightly bound to the interaction harness, tool definitions, and runtime execution environment rather than standalone model weights. Building on these diagnostic foundations, we ground our approach in an empirical analysis of over 140,000 execution trajectories.

Existing work has developed taxonomies, process representations, and failure diagnostics for coding-agent trajectories. We complement these efforts with a large-scale cross-stack analysis, using outcome-matched comparisons to isolate workflow differences from performance and rollout variability, and testing whether execution profiles remain stable across disjoint repositories and programming languages.
\section{Study Design and Behavioral Measurement}
\label{sec:study_design}

Our study analyzes execution trajectories of coding agents evaluated on tasks from the SWE-rebench Leaderboard. \footnote{Publicly released on Hugging Face by the benchmark authors: \href{https://huggingface.co/datasets/nebius/SWE-rebench-leaderboard}{SWE-rebench Leaderboard}.}  The dataset comprises 860 Python tasks across 413 repositories (spanning January 2025 to May 2026) and a multilingual set of 111 tasks across Go, Java, Python, Rust, and TypeScript. In total, we analyze a large-scale corpus of 143,209 complete execution trajectories (134,329 on Python and 8,880 on multilingual tasks) spanning 125 distinct agent configurations built upon 117 unique base models. \footnote{Code, the deterministic parser, and the analysis pipeline: \href{https://github.com/SWE-rebench/trajectory-analyzer}{SWE-rebench/trajectory-analyzer} on GitHub. Trajectories: \href{https://huggingface.co/datasets/nebius/SWE-rebench-leaderboard-trajectories}{SWE-rebench Leaderboard Trajectories} on Hugging Face.} For each task, the benchmark records the initial repository state, issue description, and an isolated test environment.

To support our empirical evaluation, we structure these data into dedicated analytical sets: the full Python corpus of 111 configurations across 413 repositories and the multilingual set of 16 frontier configurations (which share two configurations, yielding 125 distinct configurations in their union) are used to evaluate cross-repository and cross-language behavioral retrieval (Section~\ref{sec:fingerprints}), while a fixed subset of 110 spring 2026 Python tasks (March--May 2026) with multiple rollouts per task serves as a controlled panel for the outcome-matched pair design, the split-half null control, and the generational and scaffold comparisons in Section~\ref{sec:behavioral_shifts}.

\begin{figure*}[!t]
    \centering
    \includegraphics[width=\textwidth]{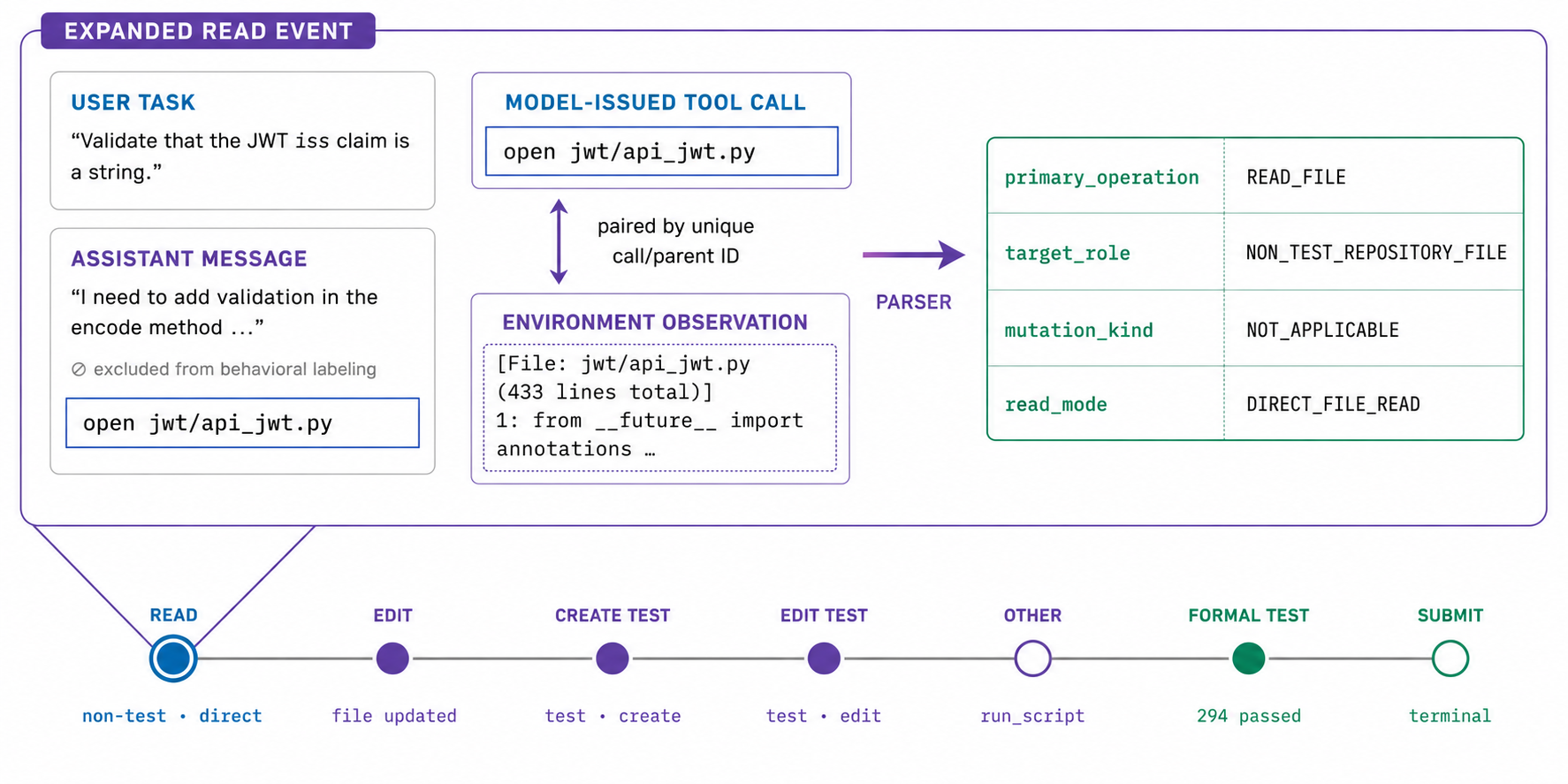}
    \caption{Automated action classification pipeline. The rule-based parser pairs each tool call with its execution output and classifies it across four core attributes: primary operation, target file role, mutation type, and read mode.}
    \label{fig:parser_pic}
\end{figure*}

Throughout this work, an \textbf{agent stack} refers to the complete deployed system: the base model, scaffolding harness (e.g., SWE-agent-like, Claude Code, Codex), reasoning-effort level, and API decoding parameters. Deploying the same base model under different scaffolds is treated as a distinct stack.

Each trajectory consists of structured messages, tool calls, and environment observations mapped via unique identifiers. We do not inspect reasoning traces; our findings characterize observable execution workflows, not internal reasoning processes.

\subsection{Action Classification}
The basic unit of analysis is an individual tool call. We implement a rule-based parser that categorizes shell commands, scripts, and API tool invocations via pattern matching into four attributes:

~\textbf{Primary operation}, classifying the core action from ten categories: exploration, search, file reading, file mutation, formal test execution, workspace inspection (\texttt{git diff}, \texttt{git status}, \texttt{git show}), package builds, other operations, no repository operation, or unknown;
~\textbf{Target file role}, classifying paths via path-based heuristics (such as \texttt{tests/}, \texttt{test\_*}, \texttt{*\_spec}, \texttt{conftest}) into non-test repository files, conventional test files, mixed targets, unresolved destinations, or not applicable;
~\textbf{Mutation type}, classifying modifications into create, edit, delete, or mixed mutations; and
~\textbf{Read mode}, distinguishing direct file reads from viewport continuations (with \texttt{read\_scope} separately capturing full-file vs.\ bounded line ranges).
Total action counts include all attempted commands (including malformed and rejected tool calls), while action proportions are calculated over executed semantic operations.

\begin{figure}[!t]
  \centering
  \includegraphics[width=\linewidth]{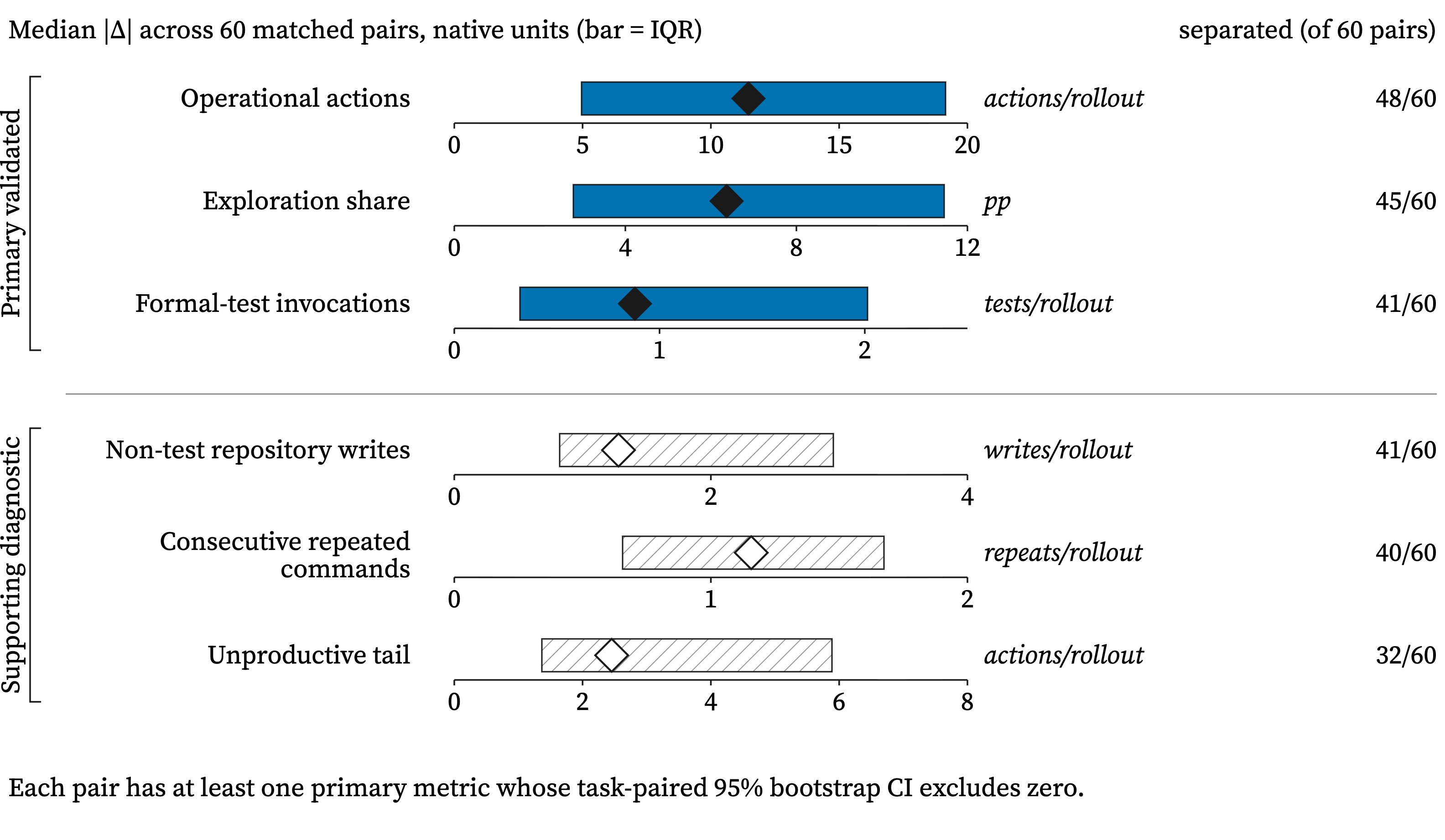}
  \caption{\textbf{Workflow differences across outcome-matched agent pairs.}
  Points indicate the median task-level absolute difference across 60 scaffold-matched agent configuration pairs ($|\Delta_{\mathrm{solve}}| \le 1.0\,\mathrm{pp}$ on $\ge 50$ shared tasks), with bars showing the interquartile range (IQR).
  The right column reports the count of pairs whose task-paired 95\% bootstrap confidence interval excludes zero.
  Top: primary workflow metrics. Bottom: supporting metrics.}
  \label{fig:workflow_divergence}
\end{figure}

\subsection{Core Workflow Metrics}

We evaluate six workflow metrics in their native units:
\begin{itemize}
    \item Operational actions: Total attempted nonterminal tool calls per rollout, including malformed or rejected attempts.
    \item Exploration share: Proportion of executed semantic actions spent on repository exploration, searching, and file reading.
    \item Non-test repository writes: Executed file modifications targeting non-test repository paths with no observed failure (excluding events with nonzero exit codes, failed tests, or execution errors).
    \item Formal-test invocations: Count of recognized test-framework invocations contained within executed events (which may exceed one in compound events).
    \item Consecutive repeated failures: Count of strictly adjacent failed attempts sharing the same canonical signature (ordered operations, normalized command, and sorted paths; a subsequent success is not required).
    \item Unproductive tail: Attempted nonterminal actions after the agent's final productive event (either a counted non-test write or a passing test).
\end{itemize}

We aggregate metrics across five runs per task for outcome-matched and case-study panels (and two rollouts per task in the configured-effort panel) before computing task averages. See Appendix~\ref{app:metric_definitions}.

\subsection{Parser Audit via Multi-Model Consensus}
\label{sec:parser_validation}

To verify parser predictions without manual labeling, we evaluate agreement against a consensus panel of six frontier LLMs (\texttt{claude-opus-5}, \texttt{gpt-5.6-sol}, \texttt{gemini-3.7-flash}, \texttt{Kimi-K3}, \texttt{DeepSeek-V4-Flash-0731}, and \texttt{MiniMax-M3}) across a stratified sample of 500 actions (Table~\ref{tab:parser_validation}). For each field, we establish a reference label only when at least 5 of the 6 models agree on the exact category. See Appendix~\ref{app:parser_validation_setup}.

\subsection{Outcome-Matched Pair Design and Null Control}
To test whether agents with equal performance follow different workflows, we compare pairs of agent configurations that share the same scaffold but differ in their base model, requiring an aggregate solve-rate difference $\vert{}\Delta_{\mathrm{solve}}\vert{} \le 1.0$ percentage points across at least 50 shared tasks. To distinguish between-stack behavioral differences from run-to-run variation across repeated rollouts, we establish a split-half null control. For each stack, we repeatedly partition its multiple rollouts into two equal, independent halves (four of the five rollouts per task; see Appendix~\ref{app:split_half_null}) across 100 random split seeds. Measuring the divergence between halves of the exact same configuration yields an empirical baseline for same-stack run-to-run variation, serving as an A/A test. Differences between distinct model pairs are then compared with this empirical null baseline (Section~\ref{sec:outcome_matched}).
\section{Empirical Results}

We organize the empirical analysis around three objectives: (1) quantify workflow differences between agents with closely matched solve rates; (2) characterize behavioral shifts associated with model choice, scaffolding, and reasoning effort; and (3) evaluate whether aggregate behavioral profiles remain discriminative across disjoint repositories and programming languages.

\subsection{Workflow Divergence Under Matched Success Rates}
\label{sec:outcome_matched}

\paragraph{Reliability of Action Classification.}
Auditing our automated parser against a consensus panel of six frontier LLMs on the stratified 500-action sample yields 82.4\% agreement on primary operations and 96.6\% on read modes, with lower agreement on target file roles and mutation types (Table~\ref{tab:parser_validation}). We therefore treat metrics that depend strongly on path resolution or mutation classification as diagnostic proxies.

\begin{table}[ht]
\centering
\small
\caption{\textbf{Multi-model consensus audit of the rule-based parser.} Reference labels require agreement among at least 5 of 6 frontier LLMs across the 500-action sample.}
\label{tab:parser_validation}
\begin{tabular}{lccc}
\toprule
\textbf{Target Attribute} & \textbf{Consensus Rate} & \textbf{Agreement with Consensus} & \textbf{Macro-F1} \\
\midrule
Primary operation                          & 94.0\% & 82.4\%  & 88.8\% \\
Target file role                           & 96.6\% & 77.0\%  & 57.8\% \\
Mutation type                              & 98.2\% & 68.8\%  & 73.7\% \\
Read mode                                  & 99.0\% & 96.6\%  & 98.4\% \\
Ordered operations                         & 87.8\% & 86.0\%  & 71.6\% \\
Read scope                                 & 100.0\% & 100.0\% & 100.0\% \\
\midrule
\textbf{Exact match (all 4 shared fields)} & 91.4\% & \textbf{77.5\%} & --- \\
\bottomrule
\end{tabular}
\end{table}

\begin{table}[t]
\centering
\small
\setlength{\tabcolsep}{5pt}
\caption{\textbf{Behavioral separation in matched pairs vs.\ null baseline.} Nominal proportion of pairs with a task-paired 95\% interval excluding 0 on $\ge k$ of the six workflow metrics.}
\label{tab:null_control}
\begin{tabular}{@{}lcc@{}}
\toprule
Separated metrics & \makecell{Split-half\\null} & \makecell{Matched pairs\\($|\Delta|\le 1$\,pp)} \\
\midrule
$\ge 1$ metric  & 30.53\% & 100.0\% \\
$\ge 2$ metrics & 6.40\%  & 100.0\% \\
$\ge 3$ metrics & 0.781\% & 96.7\%  \\
\bottomrule
\end{tabular}
\end{table}

\paragraph{Closely matched solve rates conceal behavioral divergence.}
We evaluate all 60 unique pairs of agent configurations from our corpus that share the same scaffolding harness and achieve similar solve rates across at least 50 shared tasks (median $|\Delta_{\text{solve}}| = 0.40\text{ pp}$, with a maximum difference $|\Delta_{\text{solve}}| \le 1.0\text{ pp}$). Every pair (60/60) separates on at least two of the six workflow metrics under the task-paired 95\% bootstrap criterion, and 58/60 (96.7\%) separate on at least three (e.g., action volume, exploration ratio, or test frequency). Restricting the comparison to the three primary workflow metrics (operational actions, exploration share, and formal-test invocations), every single pair separates on at least one of these three dimensions. While matched aggregate solve rates can accommodate different subsets of solved tasks across the shared panel, selected case studies further show that large workflow differences can persist even on tasks where both agents succeed in all five rollouts (as detailed in Appendix~\ref{app:case_studies}).

\paragraph{Calibration via Null Control.}
Our split-half null baseline confirms that this separation is not an artifact of run-to-run variance: while individual metrics occasionally diverge by chance, simultaneous separation on $\ge 3$ of the six workflow metrics occurs in only 0.78\% of identical-stack null splits, compared to 96.7\% across real outcome-matched pairs (Table~\ref{tab:null_control}; Appendix~\ref{app:split_half_null}).

\FloatBarrier

\subsection{Behavioral Shifts: Scaffolds, Reasoning Effort, and Model Evolution}
\label{sec:behavioral_shifts}

We examine how agent behavior shifts across model families, configured reasoning budgets, and external harnesses (Table~\ref{tab:generational_effort_comparison} and Table~\ref{tab:scaffold}), tracking specific execution mechanisms (such as \texttt{sed -n} slicing, inline Python scripts, and quiet test flags) alongside native scaffold commands.

\paragraph{Interaction Patterns across Model Generations.}
Across successive model releases under the generic harness (evaluated at High effort for Claude Opus and Medium effort for GPT), agents exhibit shifts in command structure (Table~\ref{tab:generational_effort_comparison}). In the Claude Opus lineage, later models rely far more heavily on targeted line-range slicing via \texttt{sed -n} (rising from 6.9\% in Opus 4.6 to 57.7\% in Opus 5) and adopt quiet test flags (\texttt{pytest -q}, rising from 22.7\% to 99.8\%). Similarly, newer GPT models compress their action sequences, reducing total tool operations from 24.4 in GPT-5.4 to 12.8 in GPT-5.6 Sol while routing 72.1\% of recognized edit events through inline Python scripts. These patterns produce more compressed interaction traces.

\begin{table}[!htbp]
\centering
\small
\caption{\textbf{Generational evolution and reasoning-effort scaling across workflow metrics} (110 shared tasks).}
\label{tab:generational_effort_comparison}
\setlength{\tabcolsep}{3.6pt}
\begin{tabular}{l | cccc | cccc | cccc}
\toprule
 & \multicolumn{4}{c|}{\textbf{Claude Opus (SWE)}} & \multicolumn{4}{c|}{\textbf{GPT (SWE)}} & \multicolumn{4}{c}{\textbf{Opus 4.8 (CC)}} \\
\textbf{Metric} & \textbf{4.6} & \textbf{4.7} & \textbf{4.8} & \textbf{5} & \textbf{5.4} & \textbf{5.5 M} & \textbf{5.5 xH} & \textbf{5.6 Sol} & \textbf{Low} & \textbf{Med} & \textbf{High} & \textbf{xHigh} \\
\midrule
Total actions & 47.2 & 42.8 & 33.2 & 47.6 & 24.4 & 24.2 & 43.4 & 12.8 & 31.3 & 35.5 & 39.8 & 54.4 \\
File-read share (\%) & 32.3 & 28.4 & 25.8 & 27.8 & 29.2 & 26.4 & 27.5 & 30.3 & 23.2 & 24.3 & 23.9 & 25.2 \\
File-mutation share (\%) & 12.7 & 15.9 & 16.1 & 12.9 & 14.9 & 15.8 & 11.2 & 12.3 & 19.9 & 18.0 & 17.1 & 15.4 \\
Workspace-inspection share (\%) & 4.4 & 3.1 & 3.9 & 4.6 & 6.9 & 8.5 & 8.7 & 11.4 & 0.7 & 0.9 & 1.3 & 2.2 \\
\texttt{sed -n} reads (\% of reads) & 6.9 & 58.8 & 61.0 & 57.7 & 53.6 & 87.4 & 79.9 & 76.4 & 3.6 & 3.3 & 2.8 & 2.4 \\
Python edits (\% of edits) & 0.3 & 4.6 & 4.3 & 7.7 & 55.1 & 69.0 & 72.7 & 72.1 & 0.1 & 0.1 & 0.3 & 0.2 \\
\texttt{pytest -q} (\% of tests) & 22.7 & 71.7 & 96.2 & 99.8 & 97.6 & 100.0 & 99.5 & 99.4 & 98.0 & 98.1 & 97.7 & 89.9 \\
\bottomrule
\end{tabular}
\end{table}

\paragraph{Higher Reasoning Effort Increases Exploration and Workspace Inspection.}
Holding model weights and scaffolding constant, increasing configured reasoning effort increases the number of external actions: raising GPT-5.5 effort from Medium to xHigh increases total tool actions from 24.2 to 43.4 under the generic harness, while Claude Opus 4.8 under Claude Code scales from 31.3 actions at Low to 54.4 at xHigh (Table~\ref{tab:generational_effort_comparison}). 

\begin{table}[t]
\centering
\small
\setlength{\tabcolsep}{8pt}
\caption{\textbf{Code editing and reading mechanisms across scaffolding harnesses.} Values show shares (\%) of recognized edit and read events.}
\label{tab:scaffold}
\begin{tabular}{@{}llrrrr@{}}
\toprule
Model & Scaffold & \makecell[r]{Native\\Edit} & \makecell[r]{Shell\\Redirection} & \makecell[r]{Python\\Script} & \makecell[r]{Native\\Read} \\
\midrule
Opus 4.7 & SWE-agent-like & 37.0 & 24.4 & 4.6 & 5.5 \\
Opus 4.7 & Claude Code & 92.4 & 5.4 & 0.2 & 90.5 \\
GPT-5.5 & SWE-agent-like & 0.5 & 14.9 & 69.0 & 0.0 \\
GPT-5.5 & Codex & 30.9 & 7.1 & 54.6 & 0.0 \\
\bottomrule
\end{tabular}
\end{table}

These additional actions are not concentrated in unproductive repetition: the Low-to-xHigh unproductive-tail contrast for Opus 4.8 in Claude Code is $-0.50$ actions. Instead, they go toward wider exploration: exploration and read actions increase by 12.71, and the proportion of rollouts that attempted a workspace-inspection operation after the final counted write rises from 12.7\% at Low to 66.8\% at xHigh. Read proportions remain comparatively stable across effort levels ($\sim$23\%--25\%).

\begin{table}[!ht]
\centering
\small
\caption{\textbf{Exact deployed-stack retrieval across unseen environments.}
Evaluating cross-repository transfer (111 stacks, chance $= 0.9\%$) with an 8-feature profile and cross-language transfer (16 stacks across 5 languages, chance $= 6.25\%$) with a separate 12-feature profile. Row labels denote within-panel representations, not a shared feature schema.}
\label{tab:retrieval_summary}
\setlength{\tabcolsep}{5pt}
\begin{tabular}{l cc | cc}
\toprule
 & \multicolumn{2}{c|}{\textbf{Cross-Repository (111 stacks)}} & \multicolumn{2}{c}{\textbf{Multilingual (16 stacks)}} \\
\textbf{Feature Representation} & \textbf{Top-1} & \textbf{Top-3} & \textbf{Top-1} & \textbf{Top-3} \\
\midrule
\textbf{Full behavioral profile} & \textbf{85.1\%} & \textbf{96.8\%} & \textbf{90.3\%} & \textbf{98.8\%} \\
Action-coordinate subset & 62.6\% & 86.5\% & 87.5\% & 98.4\% \\
\bottomrule
\end{tabular}
\end{table}

\paragraph{Scaffolding Strongly Shapes How Models Edit and Read Code.}
As shown in Table~\ref{tab:scaffold}, the external harness strongly influences how models interact with the environment. Under generic bash harnesses (SWE-agent-like), models rely heavily on shell-level commands such as bash stream redirections (\texttt{cat << 'EOF'}) and inline Python rewriting scripts. In the displayed Claude Code stack, 92.4\% of recognized edits and 90.5\% of recognized reads use native APIs, while the Codex stack routes 30.9\% of recognized edits through native editing.

\FloatBarrier

\subsection{Behavioral Profile Retrieval Across Repositories and Languages}
\label{sec:fingerprints}

We evaluate whether aggregate behavioral profiles remain stable across environments and are sufficiently distinctive to retrieve the corresponding agent stack on disjoint repository and language splits.

\paragraph{Feature Representation and Retrieval Protocol.}
For cross-repository retrieval, we construct an 8-dimensional profile summarizing execution style: four normalized action coordinates (exploration, file reading, non-test repository writes, and formal-test invocations) alongside four sequential verification features (normalized entropy, tool-switch rate, post-write workspace inspection, and unverified termination; see Appendix~\ref{app:fingerprint_metric_dictionary}). These vectors strictly exclude solve outcomes, total action counts, token volume, and native tool names.

To test cross-repository transfer, we partition the 413 Python repositories into two disjoint halves (207 vs.\ 206). Evaluating 111 stacks via cosine nearest-neighbor retrieval across splits (222 queries), the full behavioral profile achieves 85.1\% Top-1 recall and 96.8\% Top-3 recall against a 0.9\% random baseline (Table~\ref{tab:retrieval_summary}), showing that profiles remain discriminative across these repository splits. However, repository-disjoint splits alone do not control for differences in task composition across stacks. These differences may contribute to retrieval performance, so this result does not establish task-independent behavioral signatures.

\paragraph{Cross-Language Generalization.}
We apply the same cosine-retrieval protocol across a balanced panel of 16 agent configurations evaluated on Go, Java, Python, Rust, and TypeScript (111 tasks, 8,880 trajectories). Across all ten 3-vs-2 language splits, cross-language profile matching achieves 90.3\% Top-1 recall and 98.8\% Top-3 recall (against a 6.25\% random chance baseline). This indicates that operational signatures remain distinct across substantially different programming languages. Because action composition alone already reaches 87.5\% Top-1 in this panel (Table~\ref{tab:retrieval_summary}), we interpret these profiles as low-level operational signatures of the deployed stack rather than as evidence about higher-level problem-solving strategies or internal reasoning. See Appendix~\ref{app:retrieval_details} for feature ablations and scaling checks.

\section{Discussion and Limitations}

Our results support three main conclusions. First, aggregate solve rate provides an incomplete description of agent behavior: similarly performing stacks can exhibit substantially different observable workflows. Second, observable behavior reflects the entire agent stack, with systematic shifts associated with the underlying model, scaffold, and configured reasoning effort. Third, behavioral profiles remain discriminative across the evaluated repository and language splits, although task-composition differences may contribute to the cross-repository result. Together, these findings motivate treating execution behavior as a complementary dimension of agent evaluation alongside task success.

This perspective becomes relevant as agents operate over longer horizons and perform more actions without direct human intervention. Final outcomes reveal whether an agent succeeded, but not how it reached that outcome or what behaviors emerged along the way. Execution trajectories therefore provide process-level evidence for diagnosing and comparing agent systems. For example, trajectory audits could surface reward hacking or superficial hardcoding; behavioral profiles could reveal shared or changing operational patterns across models and scaffolds; and structured trajectory signals could provide supervision for process-based reward models and reinforcement-learning objectives. Relating behavioral profiles to token cost and latency would further show which workflows are preferable in practice. We leave these applications to future work. However, behavioral profiles are heuristic summaries rather than semantic ground truth and may be over-interpreted as evidence about latent reasoning or used for unintended system attribution.

Our study has several limitations. First, we audit our parser against a consensus of frontier LLMs rather than human annotators; agreement varies across fields (from 96.6\% for read mode to 77.0\% for target-file role and 68.8\% for mutation type), and shared systematic model biases remain possible. Second, deterministic rule-based matching captures coarse operational structure but may miss semantic context or intent in non-standard command executions. Third, our analysis focuses strictly on explicit tool calls and environment interactions, omitting internal chain-of-thought reasoning and intermediate model text. While tool calls capture concrete side effects, free-form verbal planning may contain additional context regarding model intent and latent failure modes. Fourth, all trajectories are collected in a single benchmark regime. SWE-rebench draws tasks from recent repository activity, which reduces but does not eliminate contamination, and models may also recognize the evaluation setting itself; profiles may therefore partly reflect task familiarity or benchmark-specific policies and need not transfer unchanged to real-world deployments. Fifth, matching on aggregate solve rate does not control which tasks each agent solves; matching on per-task outcomes would give a stricter comparison. Finally, our behavioral taxonomy is tailored specifically to software engineering benchmarks. While these action categories generalize well across programming languages, other domains—such as open-ended scientific research or interactive design—would require enriching the parser implementation.
\section{Conclusion}

Coding-agent performance answers whether a system succeeds, but not how it operates. Across 143,209 trajectories, we show that nearly outcome-matched agent stacks can exhibit reproducibly different workflows, that model/scaffold/effort choices systematically shift observable execution, and that aggregate behavioral profiles remain discriminative across repositories and programming languages. Together, these results motivate treating trajectory behavior as a complementary dimension of agent evaluation alongside task success. We release the parser, analysis pipeline, and an annotated trajectory subset to support further work on behavior-centered evaluation and diagnosis.

\bibliographystyle{plainnat}
\bibliography{references}

@inproceedings{jimenez2024swebench,
  title     = {{SWE}-bench: Can Language Models Resolve Real-World {GitHub} Issues?},
  author    = {Jimenez, Carlos E. and Yang, John and Wettig, Alexander and Yao, Shunyu and Pei, Kexin and Press, Ofir and Narasimhan, Karthik R.},
  booktitle = {The Twelfth International Conference on Learning Representations},
  year      = {2024},
  url       = {https://openreview.net/forum?id=VTF8yNQM66}
}

@inproceedings{yang2024sweagent,
  title     = {{SWE}-agent: Agent-Computer Interfaces Enable Automated Software Engineering},
  author    = {Yang, John and Jimenez, Carlos E. and Wettig, Alexander and Lieret, Kilian and Yao, Shunyu and Narasimhan, Karthik R. and Press, Ofir},
  booktitle = {Advances in Neural Information Processing Systems},
  volume    = {37},
  pages     = {50528--50652},
  year      = {2024}
}

@article{badertdinov2025swerebench,
  title   = {{SWE}-rebench: An Automated Pipeline for Task Collection and Decontaminated Evaluation of Software Engineering Agents},
  author  = {Badertdinov, Ibragim and Golubev, Alexander and Nekrashevich, Maksim and Shevtsov, Anton and Karasik, Simon and Andriushchenko, Andrei and Trofimova, Maria and Litvintseva, Daria and Yangel, Boris},
  journal = {arXiv preprint arXiv:2505.20411},
  year    = {2025},
  doi     = {10.48550/arXiv.2505.20411}
}

@article{kapoor2025aiagents,
  title   = {{AI} Agents That Matter},
  author  = {Kapoor, Sayash and Stroebl, Benedikt and Siegel, Zachary S. and Nadgir, Nitya and Narayanan, Arvind},
  journal = {Transactions on Machine Learning Research},
  year    = {2025},
  url     = {https://openreview.net/forum?id=Zy4uFzMviZ}
}

@article{bjarnason2026randomness,
  title   = {On Randomness in Agentic Evals},
  author  = {Bjarnason, Bjarni Haukur and Silva, Andr{\'e} and Monperrus, Martin},
  journal = {arXiv preprint arXiv:2602.07150},
  year    = {2026},
  doi     = {10.48550/arXiv.2602.07150}
}

@inproceedings{ma2024agentboard,
  title     = {{AgentBoard}: An Analytical Evaluation Board of Multi-Turn {LLM} Agents},
  author    = {Ma, Chang and Zhang, Junlei and Zhu, Zhihao and Yang, Cheng and Yang, Yujiu and Jin, Yaohui and Lan, Zhenzhong and Kong, Lingpeng and He, Junxian},
  booktitle = {Advances in Neural Information Processing Systems},
  volume    = {37},
  pages     = {74325--74362},
  year      = {2024}
}

@article{sahoo2026agentlens,
  title   = {{AgentLens}: Revealing the Lucky Pass Problem in {SWE}-Agent Evaluation},
  author  = {Sahoo, Priyam and Mittal, Gaurav and Li, Xiaomin and Ma, Shengjie and Steenhoek, Benjamin and Lin, Pingping and Hu, Yu},
  journal = {arXiv preprint arXiv:2605.12925},
  year    = {2026},
  doi     = {10.48550/arXiv.2605.12925}
}

@article{gao2026interpretagentbehavior,
  title   = {How to Interpret Agent Behavior},
  author  = {Gao, Jie and Sun, Kaiser and Huang, Jen-tse and Van Koevering, Katherine and Ji, Sijie and Huang, Heyuan and Shi, Weiyan and Lu, Zhuoran and Xiao, Ziang and Khashabi, Daniel and Dredze, Mark},
  journal = {arXiv preprint arXiv:2605.13625},
  year    = {2026},
  doi     = {10.48550/arXiv.2605.13625}
}

@article{liu2026processcentric,
  title   = {Process-Centric Analysis of Agentic Software Systems},
  author  = {Liu, Shuyang and Chen, Yang and Krishna, Rahul and Sinha, Saurabh and Ganhotra, Jatin and Jabbarvand, Reyhan},
  journal = {Proceedings of the ACM on Programming Languages},
  volume  = {10},
  number  = {OOPSLA1},
  pages   = {1961--1988},
  year    = {2026},
  doi     = {10.1145/3798271}
}

@article{oderinwale2026trajectories,
  title   = {Agent Trajectories as Programs: Fingerprinting and Programming Coding-Agent Behavior},
  author  = {Oderinwale, Hamidah},
  journal = {arXiv preprint arXiv:2606.16988},
  year    = {2026},
  doi     = {10.48550/arXiv.2606.16988}
}

@article{mehtiyev2026beyond,
  title   = {Beyond Resolution Rates: Behavioral Drivers of Coding Agent Success and Failure},
  author  = {Mehtiyev, Tural and Assun{\c{c}}{\~a}o, Wesley},
  journal = {arXiv preprint arXiv:2604.02547},
  year    = {2026},
  doi     = {10.48550/arXiv.2604.02547}
}

@article{shu2026resolverate,
  title   = {What Resolve Rate Hides: Trajectory Structure Diagnostics for Coding Agents},
  author  = {Shu, Rui and Chong, Chun Yong and Zhou, Xin and Peng, Yun and Wu, Zihan and Han, Xu and Zhuang, Zeyang and Yuan, Guowen and Wang, Yuan},
  journal = {arXiv preprint arXiv:2607.06184},
  year    = {2026},
  doi     = {10.48550/arXiv.2607.06184}
}

@article{kim2026coherencecollapse,
  title   = {Coherence Collapse: Diagnosing Why Code Agents Fail After Reaching the Right Code},
  author  = {Kim, Myeongsoo and Wang, Dingmin and Cui, Siwei and Farmahinifarahani, Farima and Zhuo, Terry Yue and Garg, Shweta and Ray, Baishakhi and Mukherjee, Rajdeep and Kumar, Varun},
  journal = {arXiv preprint arXiv:2603.24631},
  year    = {2026},
  doi     = {10.48550/arXiv.2603.24631}
}

@article{zhao2026failureprocess,
  title   = {Failure as a Process: An Anatomy of {CLI} Coding Agent Trajectories},
  author  = {Zhao, Xiangxin and Li, Han and Li, Shuaiting and Zhao, Tianyi and Barr, Earl T. and Sarro, Federica and Ye, He},
  journal = {arXiv preprint arXiv:2607.09510},
  year    = {2026},
  doi     = {10.48550/arXiv.2607.09510}
}

@article{wang2026trajaudit,
  title   = {{TrajAudit}: Automated Failure Diagnosis for Agentic Coding Systems},
  author  = {Wang, Minxing and Xie, Xiaofei and Huo, Yintong},
  journal = {arXiv preprint arXiv:2605.26563},
  year    = {2026},
  doi     = {10.48550/arXiv.2605.26563}
}

@article{raj2026modelharness,
  title   = {Model or Harness? An Interaction-Centric Taxonomy for Localizing Agent Failures},
  author  = {Raj, Harsh and Gupta, Vipul and Mahmoud, Anas and Dumitru, Razvan-Gabriel and Yi, Darvin and Sabharwal, Aakash and He, Yunzhong},
  journal = {arXiv preprint arXiv:2607.28802},
  year    = {2026},
  doi     = {10.48550/arXiv.2607.28802}
}

@inproceedings{bouzenia2025understanding,
  title     = {Understanding Software Engineering Agents: A Study of Thought--Action--Result Trajectories},
  author    = {Bouzenia, Islem and Pradel, Michael},
  booktitle = {Proceedings of the 40th IEEE/ACM International Conference on Automated Software Engineering},
  year      = {2025},
  doi       = {10.1109/ASE63991.2025.00234}
}

@article{majgaonkar2025understanding,
  title   = {Understanding Code Agent Behaviour: An Empirical Study of Success and Failure Trajectories},
  author  = {Majgaonkar, Oorja and Fei, Zhiwei and Li, Xiang and Sarro, Federica and Ye, He},
  journal = {arXiv preprint arXiv:2511.00197},
  year    = {2025},
  doi     = {10.48550/arXiv.2511.00197}
}

@inproceedings{chen2026beyondfinal,
  title     = {Beyond Final Code: A Process-Oriented Error Analysis of Software Development Agents in Real-World {GitHub} Scenarios},
  author    = {Chen, Zhi and Ma, Wei and Jiang, Lingxiao},
  booktitle = {Proceedings of the 48th IEEE/ACM International Conference on Software Engineering},
  year      = {2026},
  url       = {https://arxiv.org/abs/2503.12374}
}

@inproceedings{ceka2026traceability,
  title     = {Understanding Automated Program Repair Agents Through the Lens of Traceability: An Empirical Study},
  author    = {Ceka, Ira and Mitchell, Hailie and Pujar, Saurabh and Buratti, Luca and Ramji, Shyam and Yang, Junfeng and Kaiser, Gail and Ray, Baishakhi},
  booktitle = {Proceedings of the 35th ACM SIGSOFT International Symposium on Software Testing and Analysis},
  year      = {2026},
  url       = {https://arxiv.org/abs/2506.08311}
}

\clearpage
\appendix
\section{Behavioral Measurement and Parser Audit}
\label{app:parser_validation}

\subsection{Observable-Event Codebook and Metric Definitions}
\label{app:metric_definitions}

The parser assigns each tool invocation a \texttt{primary\_operation} from ten core categories:
\begin{itemize}
    \item exploration and search: \texttt{EXPLORE\_REPOSITORY}, \texttt{SEARCH\_REPOSITORY};
    \item reading and mutation: \texttt{READ\_FILE}, \texttt{MUTATE\_FILE};
    \item testing and inspection: \texttt{RUN\_FORMAL\_TEST}, \texttt{INSPECT\_WORKSPACE\_CHANGES};
    \item other: \texttt{PACKAGE\_OR\_BUILD}, \texttt{OTHER\_OPERATION}, \texttt{NO\_REPOSITORY\_OPERATION}, \texttt{UNKNOWN}.
\end{itemize}
For applicable operations, the parser independently tags three orthogonal attributes:
\begin{itemize}
    \item \texttt{target\_role}: distinguishes non-test repository files, conventional test files, mixed destinations, unresolved paths, or not applicable.
    \item \texttt{mutation\_kind}: classifies file modifications into create, edit, delete, mixed, or not applicable.
    \item \texttt{read\_mode}: differentiates direct file reads from viewport continuations, while \texttt{read\_scope} separately tags full-file vs.\ bounded line-range reads (\texttt{sed -n}, \texttt{head}, or API limits).
\end{itemize}

\paragraph{Action Counts and Proportions.} 
Total operational actions include all attempted nonterminal tool calls, including malformed or rejected commands. In contrast, action proportions compute the share of executed semantic operations, excluding \texttt{UNKNOWN} and \texttt{NO\_REPOSITORY\_OPERATION} events. Specifically:
\begin{itemize}
    \item \textbf{Exploration share} combines repository navigation, search commands, and file reads.
    \item \textbf{Non-test repository writes} count executed events containing a recognized mutation to at least one resolved non-test repository path with no observed failure.
    \item \textbf{Formal tests} count executed invocations recognized as project test frameworks.
\end{itemize}

\paragraph{Consecutive repeated failures.} 
This metric counts consecutive duplicate failures for events sharing the same canonical signature. For example, consider the sequence:
\[
(A, \text{fail}) \to (A, \text{fail}) \to (B, \text{fail}) \to (A, \text{success}) \to (A, \text{fail})
\]
Here, the second $A$ adds 1 repeated failure. Command $B$ starts a separate chain. The successful fourth call resets the counter for $A$, and the final failed $A$ starts a new sequence. The total count for this rollout is therefore 1. A sequence ending immediately after two failed $A$ calls also adds 1; a subsequent success is not required.

\paragraph{Unproductive Tail.} 
The unproductive tail measures attempted nonterminal tool calls after the agent's last productive step (either a counted non-test repository write without observed error or a passing formal test).
\[
\text{Read} \to \text{Counted Non-test Write} \to \text{Failed Test} \to \text{Read} \to \text{Passing Test} \to \text{Read} \to \text{Submit}
\]
The passing test marks the final productive action. The tail contains only the single subsequent read; the terminal submit action is excluded. If a rollout contains neither a counted non-test write nor a passing test, the tail includes all nonterminal tool calls.

\begin{table*}[t]
\centering
\small
\caption{Metric dictionary linking the six workflow metrics and closure diagnostics to parser fields and event rules.}
\label{tab:metric_dictionary}
\begin{tabular}{@{}>{\raggedright\arraybackslash}p{0.23\textwidth}>{\raggedright\arraybackslash}p{0.71\textwidth}@{}}
\toprule
\textbf{Metric} & \textbf{Parser field or rule} \\
\midrule
Operational actions & Attempted nonterminal structured or synthetic calls; terminal submission is excluded. \\
Exploration share & \texttt{primary\_operation} in \{\texttt{EXPLORE\_REPOSITORY}, \texttt{SEARCH\_REPOSITORY}, \texttt{READ\_FILE}\}, divided by \texttt{semantic\_denominator}. \\
File-mutation share & \texttt{primary\_operation} == \texttt{MUTATE\_FILE}, divided by \texttt{semantic\_denominator}. \\
Workspace-inspection share & \texttt{primary\_operation} == \texttt{INSPECT\_WORKSPACE\_CHANGES}, divided by \texttt{semantic\_denominator}. \\
Non-test repository writes & Executed events containing resolved non-test repository paths in \texttt{mutation\_paths} with \texttt{no\_observed\_failure} (no nonzero exit, failed test, or execution error); qualifying \texttt{MIXED} events count once. \\
Formal-test invocations & Count of recognized formal-test invocations in \texttt{contained\_operations\_ordered}; this is an invocation rate when normalized and can exceed one per semantic event. \\
Consecutive repeated failures & Strict adjacent unsuccessful events with equal canonical signatures (ordered operations, normalized command, sorted paths); missing/\texttt{UNKNOWN} breaks the chain. \\
Unproductive tail & Attempted nonterminal events after the final counted write or passing formal-test event, whichever occurs later. \\
Post-write workspace inspection & Any attempted \texttt{INSPECT\_WORKSPACE\_CHANGES} invocation occurring after the final counted non-test write anchor. \\
Termination after write & \texttt{no\_write}, \texttt{verified}, or \texttt{unverified}; verification requires a passing formal-test event after the anchor. \\
\bottomrule
\end{tabular}
\end{table*}

\subsection{Multi-Model Consensus Audit}
\label{app:parser_validation_setup}

To audit our automated parser without manual annotation, we audited its predictions against an independent consensus panel of six frontier LLMs across a stratified sample of 500 actions. We drew this sample from held-out rollouts strictly after finalizing parser rules, excluding all 1,475 development trajectories used during parser debugging.

We sampled actions across 17 stratified categories to ensure sufficient coverage of infrequent operations, such as build steps, patch diffs, and ambiguous shell commands.

We queried six independent frontier models blindly: \texttt{claude-opus-5}, \texttt{gpt-5.6-sol}, \texttt{gemini-3.7-flash}, \texttt{Kimi-K3}, \texttt{DeepSeek-V4-Flash-0731}, and \texttt{MiniMax-M3}. We concealed all parser predictions, metadata, and task outcomes from the prompts. An event receives a consensus reference label only when at least 5 of the 6 models agree on the exact category; otherwise, we mark it as unresolved.

\subsection{Fieldwise Consensus and Parser Performance}

Table~\ref{tab:parser_validation} details parser agreement with the consensus reference across classified attributes on the 500-action sample. Against the consensus panel, the parser achieves 82.4\% agreement on primary operations (88.8\% Macro-F1), 77.0\% on target file roles, 68.8\% on mutation types, and 96.6\% on read modes. Agreement on ordered operations is 86.0\%. Separately, the exact-match rate, which requires all four core fields to agree, is 77.5\%.

\paragraph{Scope and Limitations of the Audit.}
LLM-generated labels serve as a baseline, not human ground truth. This panel measures consensus across leading model families under a fixed codebook; it cannot infer latent agent intent, evaluate patch correctness, or assess omitted-action recall. Furthermore, the audit evaluates only isolated action labels and does not validate multi-step sequence metrics (such as repeated failures or the unproductive tail). 

In the consensus evaluation (Table~\ref{tab:parser_validation}), agreement on target file roles is 77.0\%. Metrics that depend strictly on file path resolution—specifically non-test repository write counts—should therefore be interpreted as diagnostic proxies rather than exhaustive filesystem assertions.

\section{Split-Half Same-Stack Null Control}
\label{app:split_half_null}

To assess whether observed workflow differences exceed run-to-run stochastic variance, we construct a split-half null baseline (an A/A test). This setup measures how much the exact same agent configuration diverges from itself across independent runs.

\paragraph{Procedure.} 
We construct pseudo-pair panels using the 120 arms from the 60 outcome-matched pairs. For each task, we randomly partition four of the five trajectories into two independent halves across 100 random split seeds, generating 12,000 split replicates (with 11,645, or 97.04\%, supporting all six metrics).

\paragraph{False Positive Rates.} 
We compute task-paired 95\% bootstrap confidence intervals between split halves of identical stacks:
\begin{itemize}
    \item \textbf{Individual metric noise:} False-separation rates remain close to nominal alpha levels: \textbf{6.30\%} for action counts, \textbf{5.97\%} for exploration share, and \textbf{6.58\%} for formal-test invocations.
    \item \textbf{Multi-metric separation:} Spurious separation on $\ge 1$, $\ge 2$, and $\ge 3$ of the six workflow metrics occurs in only 30.53\%, 6.40\%, and 0.781\% of replicates, compared to 100\%, 100\%, and 96.7\% across distinct matched pairs (Table~\ref{tab:null_control}).
\end{itemize}

This empirical baseline shows that multi-metric separation between distinct stacks is substantially larger than same-stack run-to-run variation.

\section{Agent Behavioral Signatures and Cross-Task Retrieval}
\label{app:retrieval_details}

\subsection{Behavioral Feature Representation}
\label{app:fingerprint_metric_dictionary}

To characterize each agent's problem-solving style independently of task outcomes or specific tool syntaxes, we construct an 8-dimensional behavioral vector from its executed semantic actions. It combines four normalized action coordinates (exploration, file reading, non-test repository writes, and formal-test invocations) with four sequence and verification features:
\begin{itemize}
    \item \textbf{Normalized action entropy:} Normalized Shannon entropy $H / \log(K_{\mathrm{observed}})$ across observed primary operations (imputed to 0 in the aggregation pipeline when fewer than two distinct categories are observed).
    \item \textbf{Tool-switch rate:} The frequency with which consecutive actions switch between different operational categories.
    \item \textbf{Post-write workspace inspection:} An indicator marking whether the agent attempted a workspace-inspection operation (\texttt{git diff}, \texttt{git status}, or \texttt{git show}) after the final counted non-test repository write (evaluating to false if no write occurred).
    \item \textbf{Unverified termination:} The boolean unverified-termination indicator,\footnote{Recorded as \texttt{unverified\_termination\_after\_non\_test\_repository\_write}.} marking rollouts that modified repository files without a subsequent passing formal test.
\end{itemize}
We compute these metrics per rollout and average them across runs for each agent configuration.

\subsection{Cross-Repository Retrieval Setup and Ablations}
\label{app:fingerprint_diagnostics}

To test whether behavioral signatures generalize across distinct codebases, we split our 413 Python repositories into two disjoint halves (207 vs.\ 206 repositories). For each of the 111 agent configurations, we compute its behavioral profile on Split~A and attempt to retrieve the matching profile on Split~B using cosine similarity, then repeat the process in reverse (222 total queries).

Using the full 8-dimensional profile, nearest-neighbor retrieval correctly identifies the exact deployed stack in \textbf{85.1\% of queries} (\textbf{96.8\% in Top-3}), far exceeding the 0.9\% random chance baseline (Table~\ref{tab:fingerprint_robustness}). 

\begin{table}[ht]
\centering
\small
\caption{\textbf{Exact deployed-stack retrieval across held-out repositories} (111 stacks, random chance $= 0.9\%$). Ablations use subsets of the 8-dimensional observable profile; the generic-harness subset contains 107 stacks.}
\label{tab:fingerprint_robustness}
\begin{tabular}{lcccc}
\toprule
\textbf{Feature Representation} & \textbf{Top-1} & \textbf{Top-3} & \textbf{Self Similarity} & \textbf{Nearest Non-Self} \\
\midrule
Full 8-feature profile             & \textbf{85.1\%} & \textbf{96.8\%} & 0.972 & 0.914 \\
Normalized action coordinates only$^\dagger$   & 62.6\%          & 86.5\%          & 0.971 & 0.966 \\
Sequence and verification only$^\ddagger$ & 56.8\%    & 79.7\%          & 0.960 & 0.968 \\
\midrule
Generic harness only (SWE-agent) & 85.5\%   & 97.7\%          & 0.973 & 0.914 \\
\bottomrule
\end{tabular}
\vspace{1mm}
\footnotesize
\raggedright \\
$^\dagger$\textbf{Normalized action coordinates only} includes exploration share (including reads), file-read share, non-test repository-write rate without observed failure, and contained formal-test invocation rate. \\
$^\ddagger$\textbf{Sequence and verification only} includes normalized action entropy, tool-switch rate, post-write workspace inspection after the final counted write, and the parser unverified-state compatibility indicator.
\end{table}

\paragraph{Ablations and Robustness.}
As shown in Table~\ref{tab:fingerprint_robustness}, the full profile attains 85.1\% Top-1 recall, outperforming action coordinates alone (62.6\%) and verification features alone (56.8\%). Restricting the candidate pool to 107 stacks on the generic SWE-agent harness maintains 85.5\% Top-1 accuracy (97.7\% Top-3), confirming that retrieval does not depend on scaffold-specific tool syntax.

Filtering out trajectories with unparsed commands does not alter retrieval performance (Top-1 accuracy remains between 83.8\% and 89.0\% across all parser coverage thresholds).

\subsection{Multilingual Behavioral Retrieval}
\label{app:multilingual_transfer}

To test whether behavioral signatures transfer across programming languages, we evaluate 16 agent configurations on a multilingual benchmark spanning Go, Java, Python, Rust, and TypeScript (111 tasks, 8,880 total trajectories).

\paragraph{Evaluation Protocol.}
We evaluate cross-language retrieval across all ten possible 3-vs-2 language splits. For each split, we build an agent's behavioral profile using tasks from three languages and retrieve the matching configuration against candidates evaluated on the remaining two held-out languages (and vice versa, yielding 320 directional queries). 

The multilingual representation uses a frozen 12-feature vector comprising normalized action entropy and switch rate over a ten-category general-action sequence, together with the corresponding ten action shares: repository exploration, code search, file reading, source-file creation, editing, deletion, workspace-diff inspection, other shell operations, package/build operations, and other non-shell tools. Events containing formal-test invocations are excluded from this general-action sequence, and action entropy is normalized by $\log(10)$.\footnote{Tool identity is used only to separate shell from non-shell invocations within the residual other-operation category; no tool name, task outcome, token count, or action-count magnitude enters the feature vector.}

Features are computed per rollout and averaged across runs for each task--configuration pair. Each feature is centered across configurations within each task and averaged across tasks to construct language-level profiles. Before cosine retrieval, features are scaled using the root-mean-square magnitude estimated from the query-language task-centered values, applying the same query-derived scale to candidate profiles.

\paragraph{Results and Ablations.}
Across all language splits, nearest-neighbor matching on the full 12-feature representation correctly identifies the exact deployed stack in \textbf{90.3\% of queries} (\textbf{98.8\% in Top-3}), well above the 6.25\% random chance baseline (Table~\ref{tab:multilingual_scaling}).

\begin{table}[ht]
\centering
\small
\setlength{\tabcolsep}{5pt}
\caption{\textbf{Cross-language exact-stack retrieval} (16 stacks across 5 languages, random chance $= 6.25\%$). Results pool 320 directional queries across all ten 3-vs-2 language splits using task-centered cosine retrieval.}
\label{tab:multilingual_scaling}
\setlength{\tabcolsep}{4pt}
\begin{tabular}{lccc}
\toprule
\textbf{Feature Representation} & \textbf{Top-1 Accuracy} & \textbf{Top-3 Accuracy} & \textbf{Worst-Split Top-1} \\
\midrule
Full 12-feature profile & \textbf{90.3\%} & \textbf{98.8\%} & 56.3\% \\
Action composition only (10 features) & 87.5\% & 98.4\% & 50.0\% \\
Process dynamics only (entropy + switch rate) & 36.3\% & 71.9\% & 12.5\% \\
\midrule
Common-scaffold subset (SWE-agent, 13 stacks, chance $= 7.7\%$) & 88.5\% & 98.5\% & 46.2\% \\
\bottomrule
\end{tabular}
\end{table}

\paragraph{Robustness.}
As shown in Table~\ref{tab:multilingual_scaling}, action composition alone retains 87.5\% Top-1 retrieval accuracy, while process dynamics alone reach 36.3\%: the discriminative signal is carried predominantly by action composition, with entropy and switch rate contributing a smaller complementary margin. Restricting the evaluation to the 13 stacks sharing the common SWE-agent harness yields 88.5\% Top-1 accuracy (98.5\% in Top-3) against a 7.7\% chance baseline, indicating that cross-language identification does not rely on scaffold-specific artifacts.

\section{Selected Paired Task Case Studies}
\label{app:case_studies}

To illustrate how models diverge under identical success rates, we examine three representative tasks where both Claude Opus~5 and GPT-5.6 Sol used the same generic bash harness and resolved all five independent runs. These case studies highlight three behavioral contrasts: editing interfaces, exploration granularity, and verification stopping policies.

\subsection{Case 1: Different Mutation Interfaces (Task \texttt{pypsa-1653})}
Both models solved the task across all five runs, but used different edit mechanics. Claude Opus~5 relied on multi-step shell heredoc redirections, averaging 62.4 actions, routing 48.4\% of edits through redirections, and running 12.4 formal-test invocations. In contrast, GPT-5.6 Sol routed 100\% of edits through inline Python scripts, finishing in 15.0 actions with only 3.4 test invocations.

\begin{table}[ht]
\centering
\small
\caption{\textbf{Workflow metrics for Case 1 (\texttt{pypsa-1653})}. Values are task-level means across 5 runs.}
\label{tab:case_1_metrics}
\begin{tabular}{lrr}
\toprule
\textbf{Metric} & \textbf{Claude Opus~5} & \textbf{GPT-5.6 Sol} \\
\midrule
Total tool actions & 62.4 & 15.0 \\
Python script edits (\% of edits) & 0.0\% & 100.0\% \\
Shell redirection edits (\% of edits) & 48.4\% & 0.0\% \\
Formal-test invocations & 12.4 & 3.4 \\
\bottomrule
\end{tabular}
\end{table}

\paragraph{Representative Execution Excerpts.}
Claude Opus~5 wrote files directly through bash streams before testing:
\begin{verbatim}
[11] cat << 'EOF' > source_file.py
     <REDACTED_EDIT_PAYLOAD>
     EOF
[22] python -m pytest tests/test_statistics.py -q
\end{verbatim}
GPT-5.6 Sol instead embedded programmatic file rewrites inside inline Python calls:
\begin{verbatim}
[8] python - << 'PY'
    with open("source_file.py", "r") as f: content = f.read()
    # Programmatic replacement logic
    with open("source_file.py", "w") as f: f.write(updated_content)
    PY
[9] pytest tests/test_statistics.py -q
\end{verbatim}

GPT-5.6 Sol consistently expresses complex code updates via self-contained Python transformations. This approach avoids some shell-escaping complexity and uses fewer interaction turns.

\subsection{Case 2: Granular Navigation vs. Broad Execution (Task \texttt{ytmusicapi-909})}
A lower action count does not reflect identical exploration. Claude Opus~5 averaged 67.2 iterative steps, including 25.4 exploration/read actions and 6.6 test invocations. GPT-5.6 Sol compressed its workflow into 11.6 actions (4.6 exploration/read actions and 4.6 test invocations), while both exhibited similar workspace inspection activity (5.0 vs.\ 5.4 operations).

\begin{table}[ht]
\centering
\small
\caption{\textbf{Workflow metrics for Case 2 (\texttt{ytmusicapi-909})}. Values are task-level means across 5 runs.}
\label{tab:case_2_metrics}
\begin{tabular}{lrr}
\toprule
\textbf{Metric} & \textbf{Claude Opus~5} & \textbf{GPT-5.6 Sol} \\
\midrule
Total tool actions & 67.2 & 11.6 \\
Exploration and read actions & 25.4 & 4.6 \\
Formal-test invocations & 6.6 & 4.6 \\
Workspace-inspection operations & 5.0 & 5.4 \\
\bottomrule
\end{tabular}
\end{table}

\paragraph{Representative Execution Excerpts.}
Claude Opus~5 constrained its search outputs and used quiet test flags:
\begin{verbatim}
[5]  grep -rn "def get_playlist" . | head -n 20
[25] python -m pytest tests/test_playlists.py -q
\end{verbatim}
GPT-5.6 Sol inspected broad line ranges and executed default verbose test runs:
\begin{verbatim}
[2] sed -n '1,800p' ytmusicapi/navigation.py
[9] pytest tests/test_playlists.py
\end{verbatim}

Action compression reflects different command-selection policies: one agent relies on tightly bounded exploratory queries across many turns, while the other favors wide, monolithic commands in fewer steps.

\subsection{Case 3: Extended Post-Write Verification (Task \texttt{astronomy-commons\_\_lsdb-1349})}
Agents differ substantially in their verification stopping criteria. Claude Opus~5 performed extensive post-write verification, averaging 69.8 total actions, 9.2 test invocations, 5.8 workspace inspections, and 8.8 actions after its final code write. GPT-5.6 Sol completed the task in a compact 20.2 actions, with 3.6 test invocations, 7.0 workspace inspections, and 4.4 post-write actions.

\begin{table}[ht]
\centering
\small
\caption{\textbf{Workflow metrics for Case 3 (\texttt{astronomy-commons\_\_lsdb-1349})}. Values are task-level means across 5 runs.}
\label{tab:case_3_metrics}
\begin{tabular}{lrr}
\toprule
\textbf{Metric} & \textbf{Claude Opus~5} & \textbf{GPT-5.6 Sol} \\
\midrule
Total tool actions & 69.8 & 20.2 \\
Formal-test invocations & 9.2 & 3.6 \\
Workspace-inspection operations & 5.8 & 7.0 \\
Actions after final counted write & 8.8 & 4.4 \\
\bottomrule
\end{tabular}
\end{table}

\paragraph{Representative Execution Excerpts.}
Claude Opus~5 performed multiple regression checks and inspected workspace status after editing:
\begin{verbatim}
[63] python - << 'PY' ... PY  --> (Final code modification)
[65] python -m pytest tests/test_hats.py -q  --> (Target test)
[67] python -m pytest -q  --> (Full regression suite)
[69] git status --short  --> (Workspace check)
\end{verbatim}
GPT-5.6 Sol executed a single targeted test pass and terminated immediately:
\begin{verbatim}
[16] python - << 'PY' ... PY  --> (Final code modification)
[18] pytest tests/test_hats.py -q  --> (Target test passed)
     [TERMINAL SUBMISSION]
\end{verbatim}

Matched solve rates obscure major differences in agent stopping policies: in the illustrated runs, Opus exhibits cautious multi-stage verification, whereas Sol terminates immediately upon passing a single targeted test.

\section{Configured-Effort Decomposition in Claude Code}
\label{app:opus48_effort_deep_dive}

To understand how configured reasoning effort alters observable execution, we analyze 880 trajectories across four configured effort levels of Claude Opus 4.8 in Claude Code CLI on the 110-task panel. Table~\ref{tab:opus48_decomposition} reports three selected levels; all four appear in Table~\ref{tab:generational_effort_comparison}.

\begin{table}[ht]
\centering
\small
\caption{\textbf{Trajectory structural shifts across selected configured effort levels for Claude Opus 4.8 in Claude Code} (110 common tasks, 2 rollouts per task). Formal-test volume counts contained invocations. Differences are computed before display rounding.}
\label{tab:opus48_decomposition}
\begin{tabular}{lrrrc}
\toprule
\textbf{Trajectory Dimension} & \textbf{Low} & \textbf{Medium} & \textbf{xHigh} & \textbf{Difference [95\% CI]} \\
\midrule
Total operational actions & 31.3 & 35.5 & 54.4 & $+23.0\ [16.5, 27.8]$ \\
Exploration and read actions & 14.4 & 17.8 & 27.1 & $+12.7\ [11.0, 14.5]$ \\
File-mutation events & 5.4 & 6.1 & 8.3 & $+3.0\ [2.3, 3.6]$ \\
Formal-test invocations & 4.7 & 4.5 & 5.4 & $+0.7\ [-1.8, 2.3]$ \\
\midrule
\shortstack[l]{Rollouts inspecting workspace changes\\after the final counted write (\%)} & 12.7\% & 24.1\% & 66.8\% & $+54.1\text{ pp}\ [45.9, 61.8]$ \\
Action-type switch count & 19.0 & 23.0 & 34.5 & $+15.6\ [13.7, 17.5]$ \\
Unproductive tail actions & 6.1 & 4.3 & 5.6 & $-0.5\ [-5.6, 2.5]$ \\
\bottomrule
\end{tabular}
\end{table}

\paragraph{Key Observations.}
\begin{itemize}
    \item \textbf{Exploration vs. Testing:} The 23.0-action increase from Low to xHigh is driven mainly by exploration and reading, which rise from 14.40 to 27.12 actions ($+12.71$, 95\% CI $[10.98, 14.46]$) and account for 55.2\% of the growth. File-mutation events rise from 5.38 to 8.33, while formal-test invocations stay flat ($4.65 \to 5.37$, $+0.71$, 95\% CI $[-1.75, 2.25]$).
    \item \textbf{Active Workspace Verification:} Higher reasoning effort substantially increases post-write workspace inspection (rising from 12.7\% to 66.8\% of rollouts, $+54.1$ pp) and action switching ($18.97 \to 34.52$).
    \item \textbf{Constant Unproductive Tail:} Increasing effort in Claude Code does not increase trailing unproductive actions: tail volume shifts by only $-0.50$ actions (95\% CI $[-5.60, 2.53]$), with no clear evidence of a change.
\end{itemize}

\section{Compute Resources and Infrastructure}
\label{sec:compute_resources}

\paragraph{Rollout Collection and LLM Inference.}
The evaluated trajectory rollouts were gathered using the SWE-rebench distributed evaluation infrastructure in isolated containerized environments. Model inference was conducted via external provider APIs (including OpenAI, Anthropic, Google, MiniMax, and StepFun), with most model--task configuration executed across 5 independent rollouts. For our primary comparative cohort (April--May 2026), the evaluated rollouts accounted for approximately 20.5 billion total tokens ($\approx$\$7.1k nominal API-equivalent cost based on provider accounting).

\paragraph{Offline Parsing and Statistical Analysis.}
All subsequent data processing, rule-based parsing, bootstrap resampling, null-control calibrations, and nearest-neighbor retrieval pipelines were executed entirely on a single standard multi-core CPU machine without GPU acceleration:
\begin{itemize}
    \item \textbf{Action parsing and metric aggregation:} Parsing the execution traces and extracting behavioral profiles took approximately 1--2 wall-clock hours, with a peak memory footprint under 20~GB RAM.
    \item \textbf{Split-half null baseline:} Evaluating the 100 random split seeds across all configurations completed in under 1 minute using 32 CPU worker threads.
    \item \textbf{Cross-split and multilingual retrieval:} Cosine similarity computations and nearest-neighbor rank retrieval ran in under 20 minutes (peak memory $< 5$~GB RAM).
\end{itemize}

\end{document}